\documentclass[twocolumn,twoside,slac_two]{revtex4}
\usepackage{graphicx}
\usepackage{fancyhdr}
\begin{document}

\title{Properties of Gamma Ray Bursts at different redshifts}

\author{G. Pizzichini, E. Maiorano, M. Genghini}
\affiliation{INAF/IASF Bologna, Italy}
\author{F. M\"unz}
\affiliation{DCMP, Faculty of Science, Masaryk University, Brno, Czech 
Republic}

\begin{abstract}
GRBs are now detected up to $z=8.26$~\cite{Sal-ref},\cite{Tan-ref}. We try 
to find differences, in their restframe properties, which could be related 
either to distance or to observing conditions.
\end{abstract}

\maketitle

\thispagestyle{fancy}

\section{SAMPLE SELECTION}
We try to find changes in the properties of GRBs at different redshifts 
which could be related to source evolution. We consider all the 149 events 
detected by Swift~\cite{Bar-ref},~\cite{Geh-ref} between January 26, 2005 
and July 15, 2009 for which the redshift has been at least tentatively 
measured. We use the table given at the HEASARC website\footnote{ 
http://heasarc.gsfc.nasa.gov/docs/swift/archive/grb\_table.html}.

Figure 1 shows the histogram of all GRB redshifts detected until July 15, 
2009. Data for Figure 1 were taken from the above quoted website and from 
GCN\footnote{http://gcn.gsfc.nasa.gov/gcn3\_archive.html}. We also show 
the histogram for events detected by HETE-2 and {\it Beppo}SAX, although 
the trigger criteria were not the same for all three experiments.

\section{ANALYSIS}
The Swift table gives also the BAT fluence (15-150 keV) and the BAT 
T$_{90}$ (15-350 keV). We consider both the values in the observer's frame 
and the ones converted in the restframe. The redshift in the events goes 
from 0.0331 to 8.26. Note that the lowest value of the redshift until now 
is 0.0085 for GRB980425, detected by {\it Beppo}SAX.
\begin{figure}[b!]
\centering
\includegraphics[width=8.1cm]{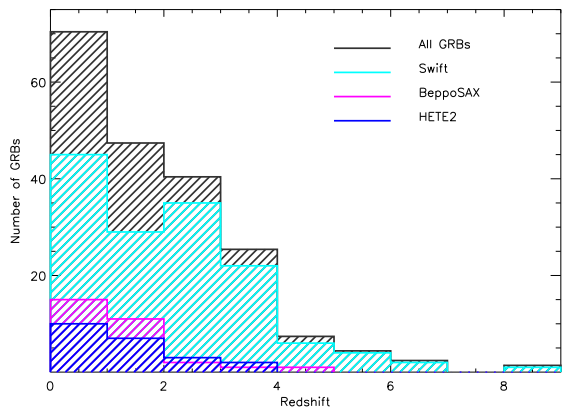}
\caption{Histogram of all the redshifts measured for GRBs until July 15, 
2009.}
\label{istog-fig}
\end{figure}

The scatter diagrams of those quantities, namely BAT T$_{90}$ and fluence, 
are shown in Figures 2 and 4 respectively, both in the observer's and in 
the restframe. For lack of some data only a total of 139 events could 
actually be used. For comparison we also show T$_{90}$ versus redshift 
(Figure 3) and fluence versus redshift (Figure 5) as obtained using the 
Gamma Ray Burst Monitor (GRBM)~\cite{Fro-ref}, on board of {\it Beppo}SAX 
(energy bands: $E>20$ keV for T$_{90}$ and 40--700 keV for fluence) and 
FREnch GAmma-ray TElescope (FREGATE)~\cite{Pel-ref}, on board of HETE-2 
(energy bands: 6--80 keV for T$_{90}$ and 2--30 keV for fluence).

As noted by Dr. Upendra Desai~\cite{Des-ref} in Figure 2 and even more in 
Figure 4 there seems to be a lack of ``low T$_{90}$--low Fluence" events 
at $z\sim1.5$ and $z\sim3$, but unfortunately the number of GRBs is still 
too small to allow us to confirm this possibility. We recall that, for 
example, in the case of quasars the periodicity of redshifts is a problem 
which has been debated for many years~\cite{Lop-ref}.

As shown in Figure~\ref{sw_z-ttrtrmed-fig}, we also tried to compensate 
for the fact that the values, taken at the same energy range in the 
observer's frame, originate from different energy ranges in the event's 
restframe. T$_{90}$ in the restframe is the observed value divided by 
$(1+z)$ and the fluence in the restframe is the observed value multiplied 
by $(1+z)$, but their respective energy intervals in the restframe are 
also multiplied by $(1+z)$. By using the Fenimore~\cite{Fen-ref} 
correlation between peak duration and energy, and assuming that it can be 
applied also to T$_{90}$, we try to take into account that, for long 
bursts, the duration normally decreases with energy.

In Figure 7 we show the scatter diagram of fluence versus T$_{90}$, again 
in the observer's and restframe for the Swift events, color coded for six 
redshift intervals.



\begin{figure*}[t!]
\includegraphics[width=10.cm]{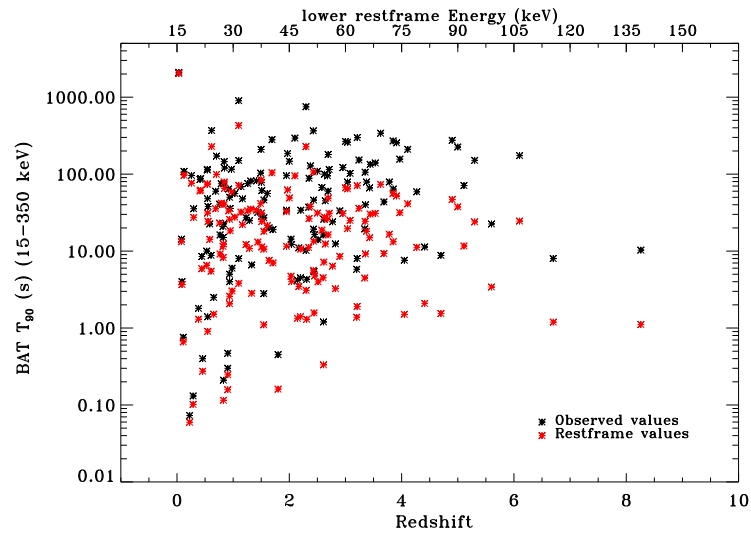}
\caption{Scatter plots of the T$_{90}$ versus burst redshift obtained for 
all the 139 events detected by Swift between January 26, 2005 and July 15, 
2009. The T$_{90}$ restframe values are, in first approximation, the 
observed ones divided by $(1+z)$, but the energy ranges must also be 
multiplied by the same value in the restframe. The top scale shows the 
lower value of the instrument energy range at that redshift.}
\label{sw_z-ttr-fig}
\end{figure*}

\begin{figure*}[ht!]
\includegraphics[width=0.45\textwidth]{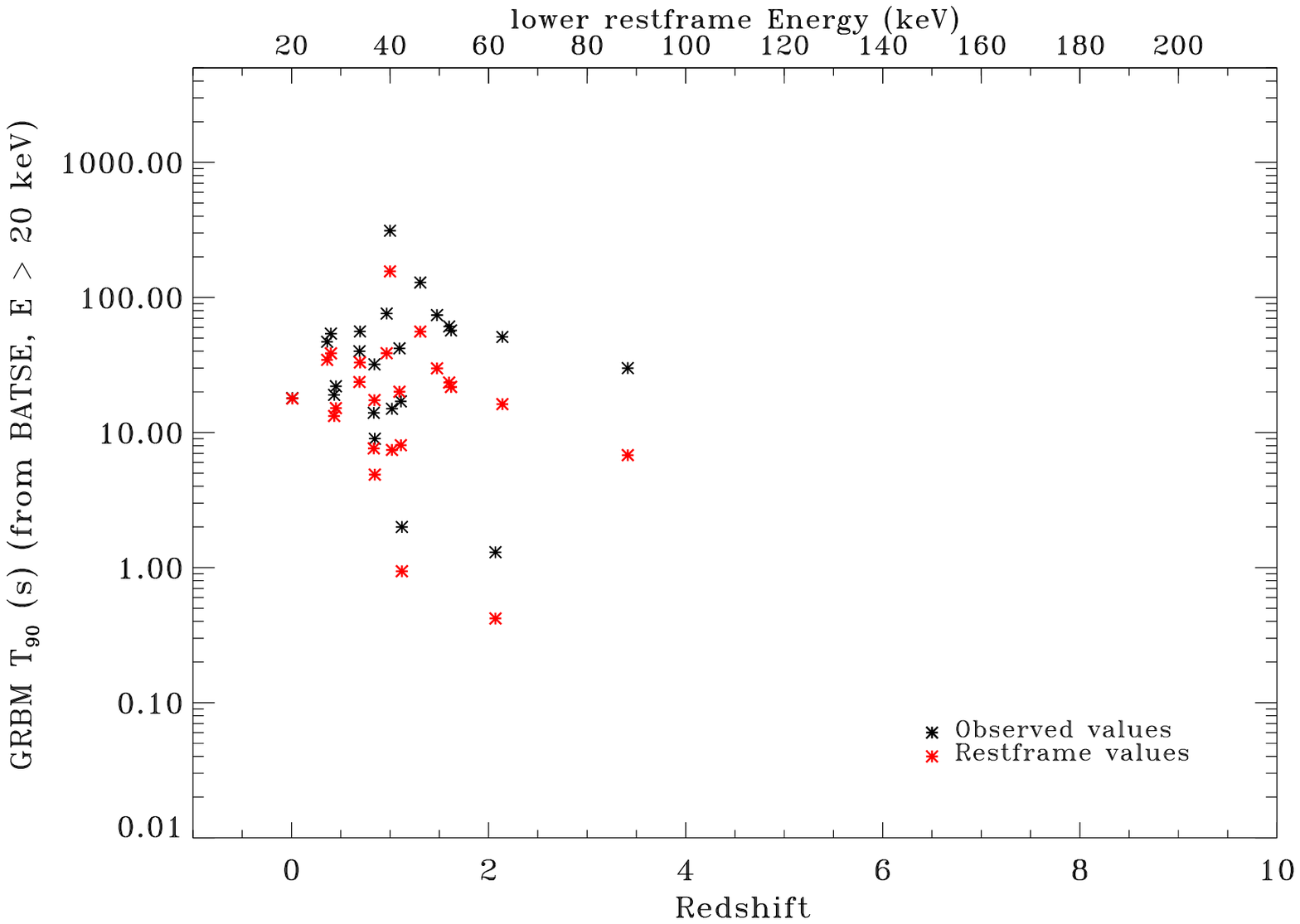}
\includegraphics[width=0.45\textwidth]{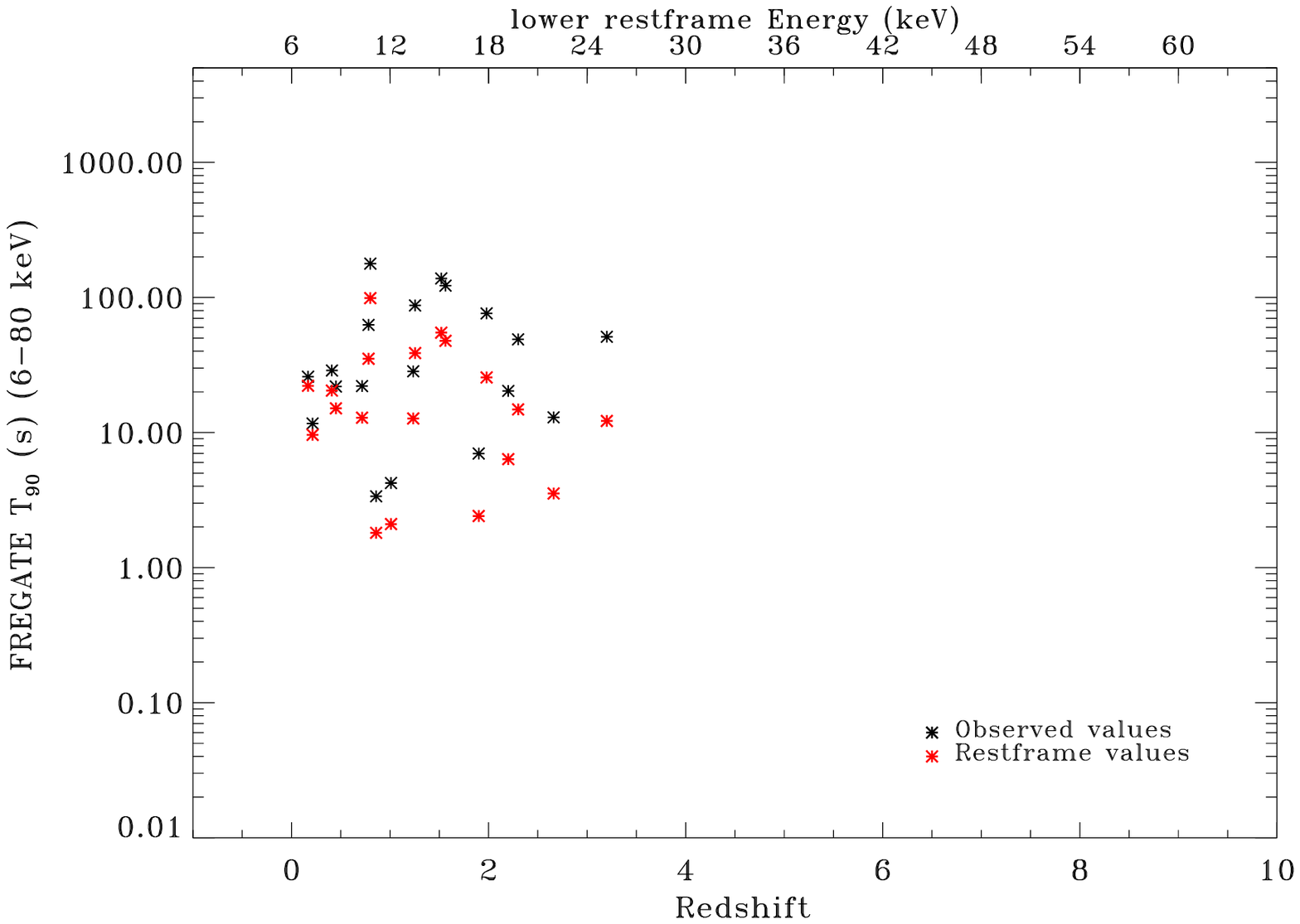}
\caption{For comparison with Figure 2 we show also the scatter plots of
T$_{90}$ versus burst redshift for {\it Beppo}SAX~\cite{Fro-ref} and
HETE-2~\cite{Pel-ref} (left and right panels respectively). The top scale
shows again the lower value of the instrument energy range at that
redshift.}
\label{sax-hete-ttr-fig}
\end{figure*}


\begin{figure*}[ht!]
\includegraphics[width=10.cm]{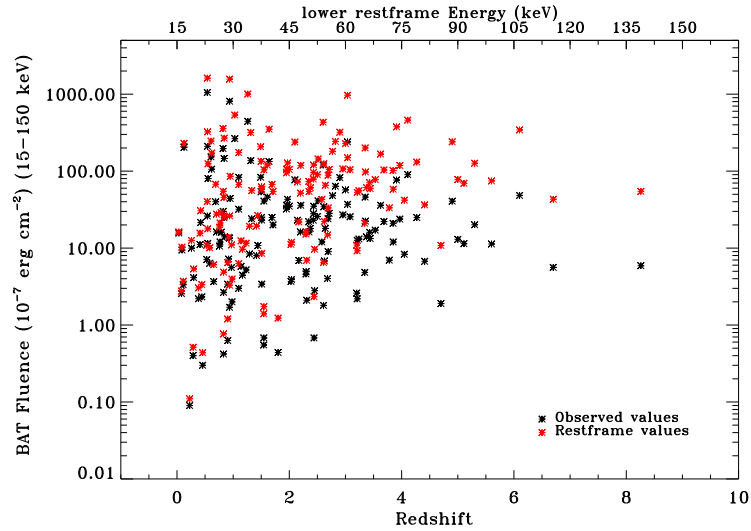}
\caption{Scatter plots of the fluence versus burst redshift obtained for 
all the 139 events detected by Swift between January 26, 2005 and July 15, 
2009. The restframe fluence values are, in first approximation, the 
observed ones simply multiplied by $(1+z)$, but, as in Figure 2 and 3, we 
must remember that the energy ranges must also be multiplied by the same 
value in the restframe. The top scale shows the lower value of the 
instrument energy range at that redshift.}
\label{sw_z-ffr-fig}
\end{figure*}

\begin{figure*}[ht!]
\includegraphics[width=0.45\textwidth]{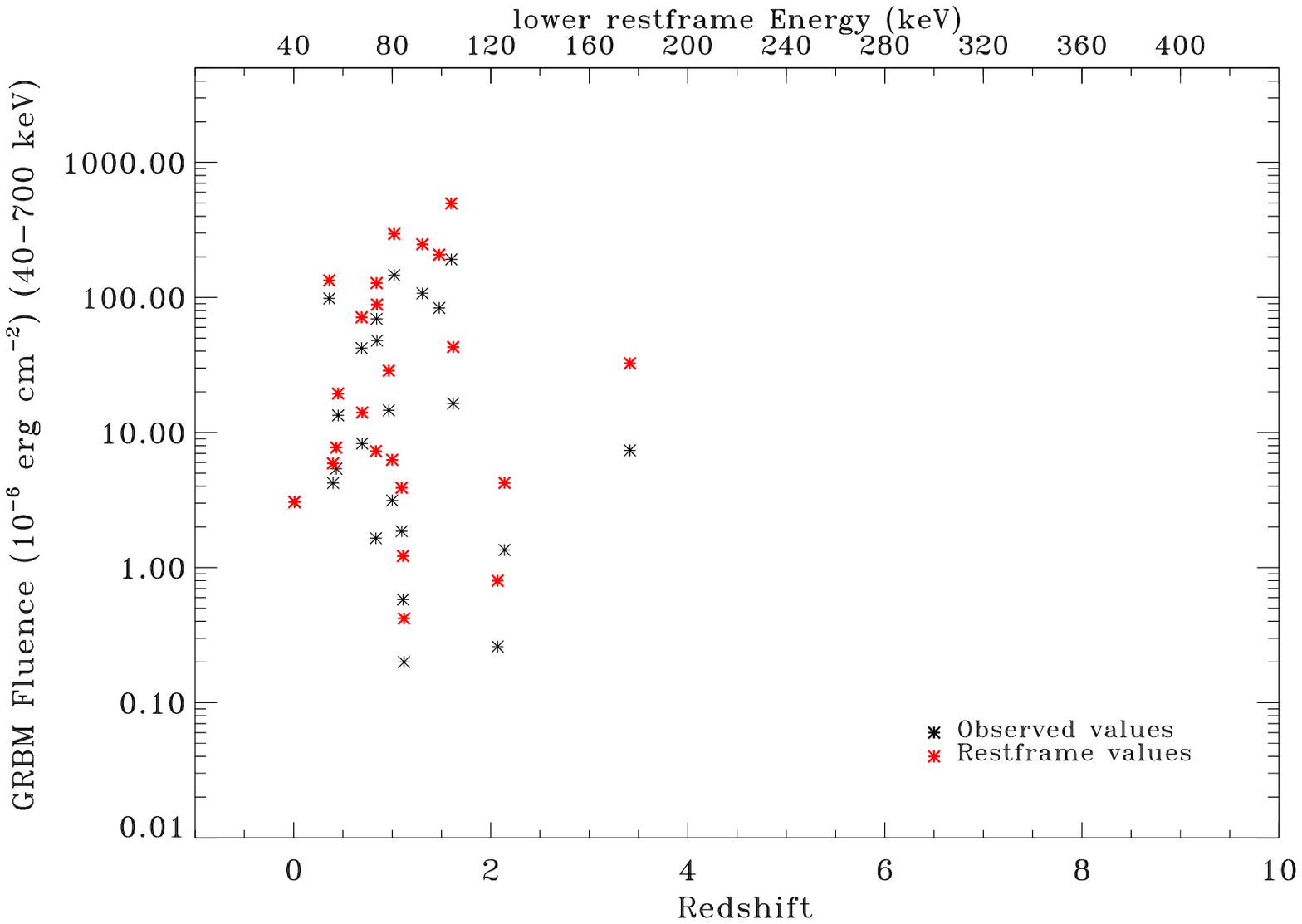}
\includegraphics[width=0.45\textwidth]{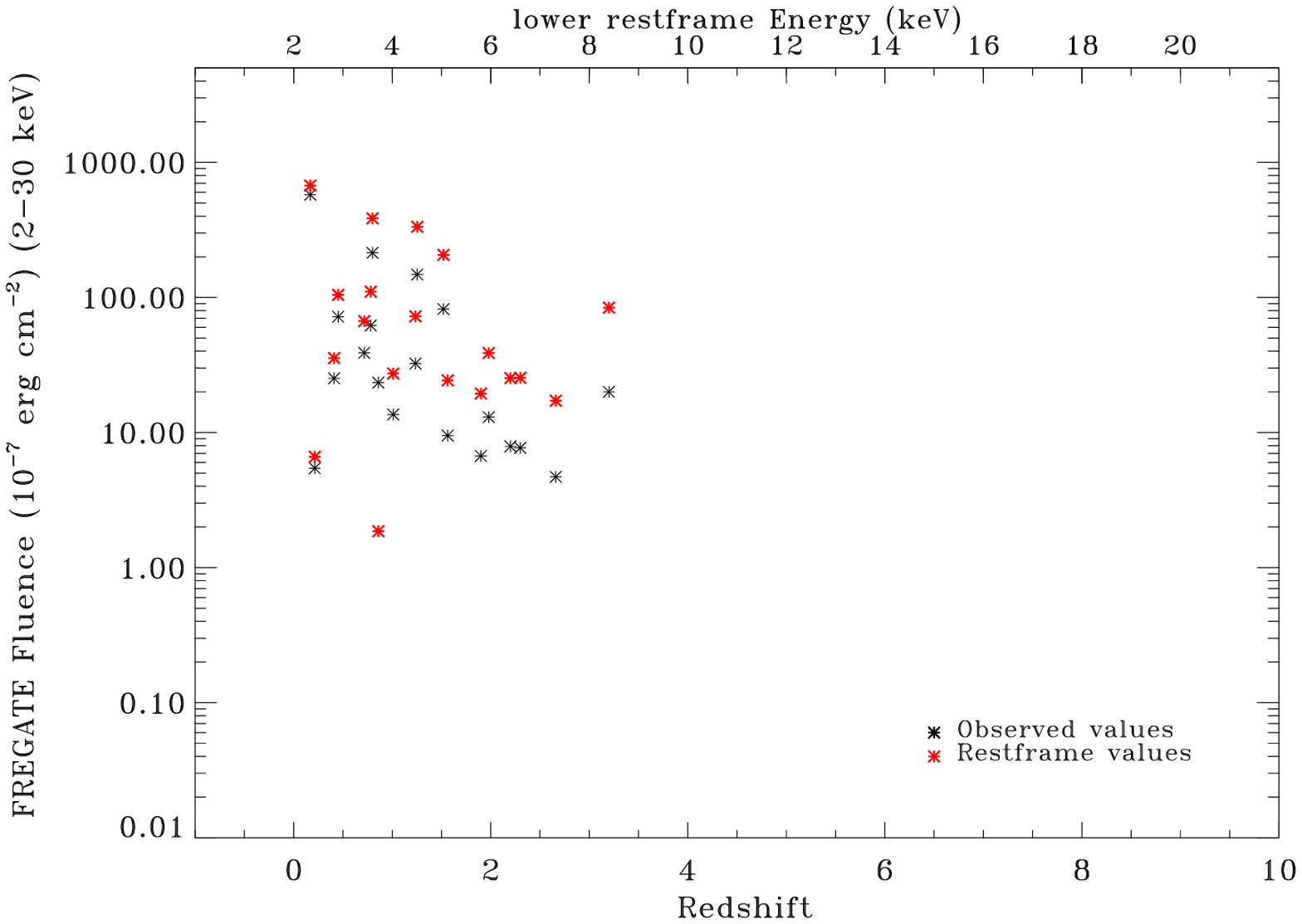}
\caption{For comparison with Figure 4 we show also the scatter plots of
the fluence versus burst redshift for {\it Beppo}SAX~\cite{Fro-ref} and
HETE-2~\cite{Pel-ref} (left and right panels respectively). The top scale
shows again the lower value of the instrument energy range at that
redshift.}
\label{sax-hete-ffr-fig}
\end{figure*}


\begin{figure*}[t]
\includegraphics[width=10.cm]{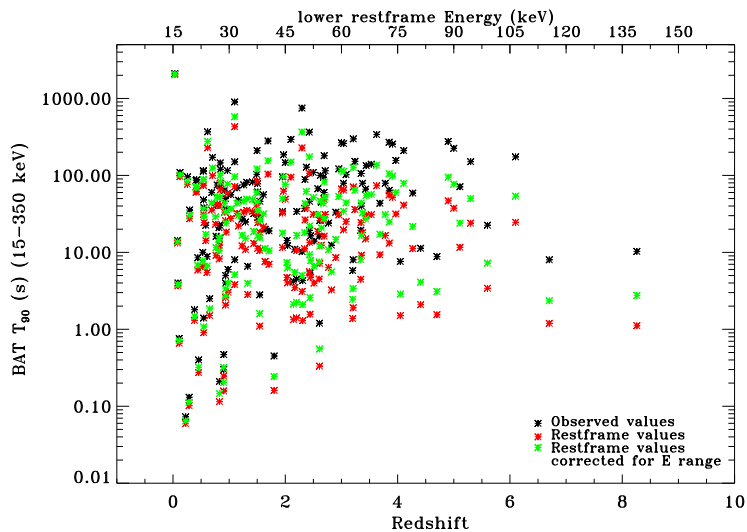}
\caption{In order to reduce T$_{90}$ in the restframe to the same energy 
range for all bursts we consider a dependence of the burst duration on the 
energy similar to the one given, for peaks, by Fenimore et 
al.~\cite{Fen-ref}. The figure shows, in green, the scatter plot of what 
would be the T$_{90}$ distribution versus redshift in that case.}
\label{sw_z-ttrtrmed-fig}
\end{figure*}

\begin{figure*}[ht!]
\includegraphics[width=0.45\textwidth]{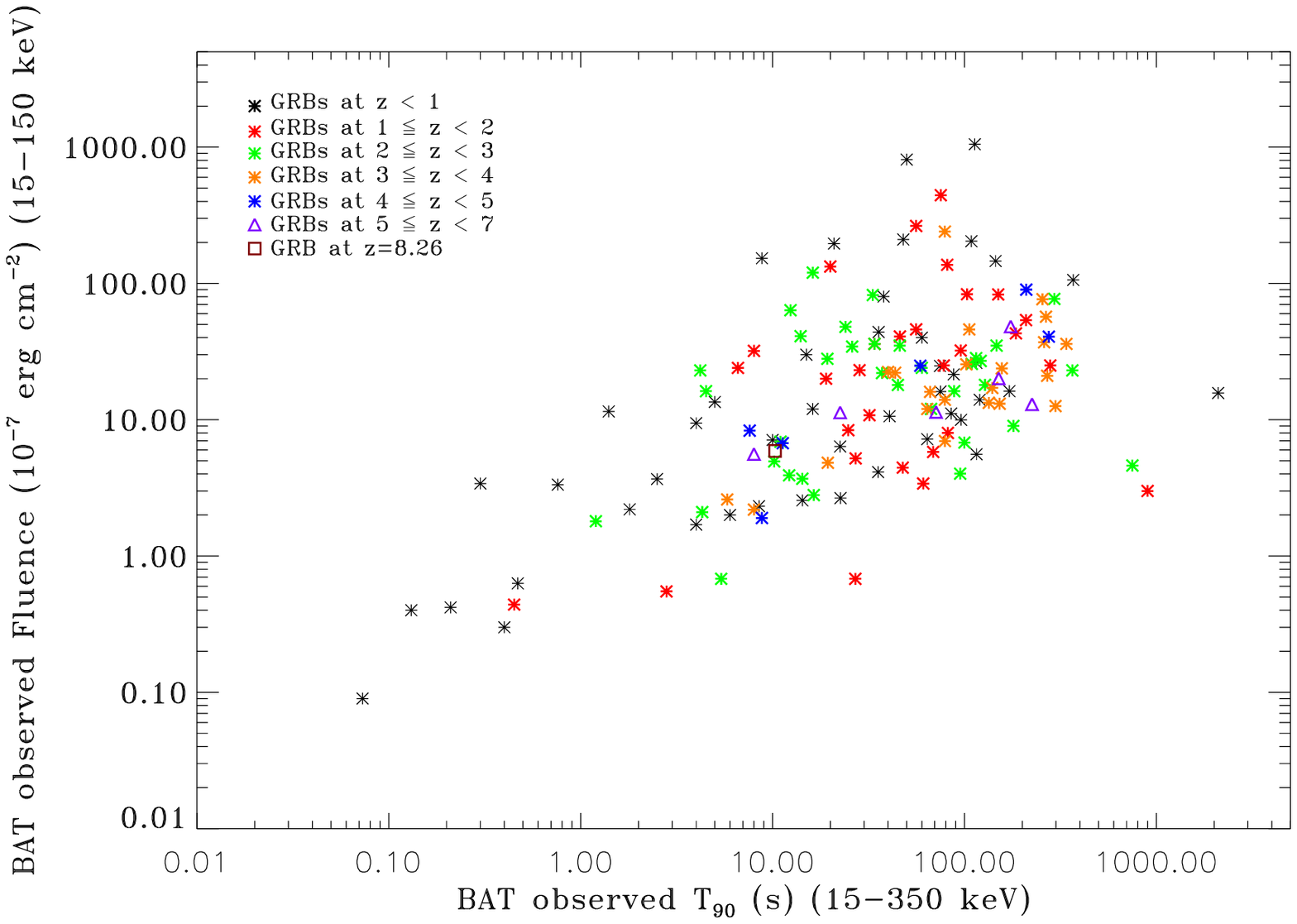}
\includegraphics[width=0.45\textwidth]{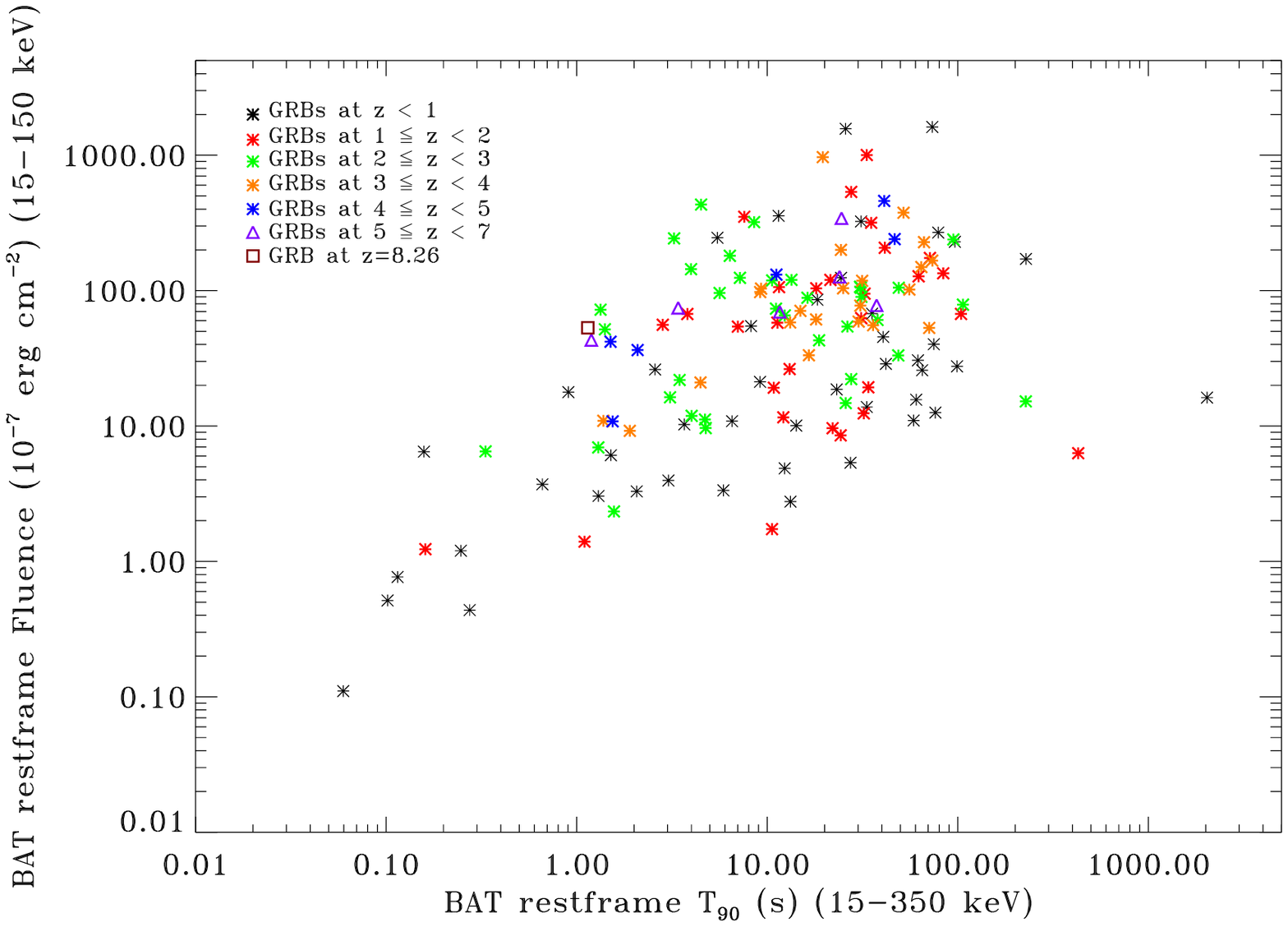}
\caption{Scatter plot of the BAT fluence (15-150 keV) in 10$^{-7}$ erg
cm$^{-2}$ versus BAT T$_{90}$ (15-350 keV), color coded for six redshift
intervals. Note that GRB090423 at $z=8.26$ (the purple empty square in
both panels), falls in the middle of the distribution in the observer's
frame. Left panel: observed values. Right panel: restframe values.
See Table~\ref{Coeff-tab} for the correlation coefficients in the six
intervals.}
\label{sw_obs-rest-fig}
\end{figure*}

\clearpage

\section{Correlation coefficients}

We also computed the fluence--T$_{90}$ correlation coefficients for the
same redshift groups shown in Figure 6 (see Table~\ref{Coeff-tab}). The
correlation for all redshifts together is $0.61$, both for observed and
restframe values, non weighted, while it is $0.46$ and $0.53$ for observed
and restframe weighted values.

\begin{table*}[]
\begin{center}
\caption{Correlation coefficients: if we use the same redshift groups as
in Figure 7, we obtain the correlation coefficients for the restframe
values of log fluence and log T$_{90}$ reported in the table. The second row
takes into account errors in fluence (errors on T$_{90}$ are not
available) using weights in the calculation inversely proportional to log
fluence error. The correlation values become lower for the first groups,
because points in the lower left hand corner carry higher fluence errors.}
\vspace{0.5cm}
\begin{tabular}{|c|c|c|c|c|c|c|}
\hline \textbf{Redshift groups} & \textbf{0--1} & \textbf{1--2} &
\textbf{2--3} & \textbf{3--4} & \textbf{4--5} & \textbf{5--7} \\
\hline \textbf{Number of GRBs} &  &  &  &  &  &  \\
\textbf{per group} & 45 & 28 & 34 & 21 & 6 & 5 \\
\hline  Coeff. neglecting &  &  &  &  &  & \\
both errors & $0.68\pm0.08$ & $0.57\pm0.13$ & $0.30\pm0.16$ & 
$0.66\pm0.12$ & $0.92\pm0.06$ & $0.81\pm0.15$ \\
\hline  Coeff. including &  &  &  &  &  & \\
fluence errors & $0.42\pm0.12$ & $0.27\pm0.18$ & $0.20\pm0.16$ & 
$0.26\pm0.20$ & $0.96\pm0.03$ & $0.81\pm0.16$ \\
\hline
\end{tabular}
\label{Coeff-tab}
\end{center}
\end{table*}

\section{CONCLUSION}

We considered 139 GRBs at different redshifts, all of them detected by the 
same experiment, Swift-BAT, hoping to find proof of evolution with $z$. 
Except for the well known two groups of ``short" and ``long" GRBs, which 
appear to be a little less well defined in the restframe, no such proof is 
evident. We can see from the plots (Figures 2 and 4) that the number of 
events obviously becomes smaller with redshift, but both T$_{90}$ and 
fluence hardly change in their average log value, where we find them also 
for the events at the largest redshifts. Thus we conclude that the 
probability of having GRB with those values is higher, even at large $z$. 
We conclude that no redshift selection or evolution can yet be inferred 
from our plots. Even GRB090423, the one detected at the 
largest $z$ until now, lies just in the middle of the distribution in the 
observer's restframe. Not 
surprisingly, it is evident from the restframe plots (Figure 7, right 
panel) that fluence increases with T$_{90}$ practically for all redshifts. 
Bursts at large redshifts have higher fluences, but we must remember that 
they originate from higher energy ranges and that detection thresholds 
favor distant intense events.

\end{document}